\documentclass[pre, twocolumn, showkeys, amsmath, amssymb]{revtex4}

\usepackage{graphicx}

\begin{document}

\title{The Effect of the SOS Response on the Mean Fitness of Unicellular Populations:  A Quasispecies Approach}

\author{Amit Kama}
\affiliation{Department of Chemistry, Ben-Gurion University of the Negev, Be'er-Sheva, Israel}
\author{Emmanuel Tannenbaum}
\email{emanuelt@bgu.ac.il}
\affiliation{Department of Chemistry, Ben-Gurion University of the Negev,
Be'er-Sheva, Israel}

\begin{abstract}

This paper develops a quasispecies model that incorporates the SOS response.  We consider a unicellular, asexually replicating population of organisms, whose genomes consist of a single, double-stranded DNA molecule, i.e. one chromosome.  We assume that repair of post-replication mismatched base-pairs occurs with probability $ \lambda $, and that the SOS response is triggered when the total number of mismatched base-pairs exceeds $ l_S $.  We further assume that the per-mismatch SOS elimination rate is characterized by a first-order rate constant $ \kappa_{SOS} $.  For a single fitness peak landscape where the master genome can sustain up to $ l $ mismatches and 
remain viable, this model is analytically solvable in the limit of infinite sequence length.  The results, which are confirmed by stochastic simulations, indicate that the SOS response does indeed confer a fitness advantage to a population, provided that it is only activated when DNA damage is so extensive that a cell will die if it does not attempt to repair its DNA.

\end{abstract}


\maketitle

\noindent
{\bf Author Summary:}  {\it Genetic repair is currently a major area of experimental research in molecular and systems biology, because the breakdown of genetic repair is believed to play a crucial role in phenomena such as the emergence of cancer and the emergence of antibiotic-resistant strains of bacteria.  As with many other research areas in biology, mathematical models can be expected to play an increasingly important role in understanding various genetic repair mechanisms in unicellular organisms.  In this vein, I have developed an analytically solvable model describing the evolutionary dynamics of a unicellular population capable of undergoing the SOS response.  The SOS response is a repair mechanism that has been receiving a considerable amount of attention recently, primarily because it is a repair mechanism that is highly error-prone, and so it is somewhat paradoxical that such a repair mechanism could confer a selective advantage.  To my knowledge, this paper is the first of its kind to mathematically model the evolutionary aspects of the SOS response, and so I believe that this work provides an initial, and much-needed, theoretical foundation for understanding the role of this repair mechanism.}

\section{Introduction}

Genetic repair is an essential component of cellular genomes.  Without mechanisms for repairing damaged and mutated DNA, genomes could not achieve sufficient information content to code for the variety and complexity of modern terrestrial life \cite{Voet}.

Genetic repair mechanisms fall into two main categories:  Those that correct base mis-pairings during the replication cycle of a cell, and those that repair mutated and damaged DNA during the growth (G) phase of the cellular life cycle \cite{Voet}.

Two important examples of the first class of repair mechanisms are DNA proofreading and mismatch repair (MMR).  DNA proofreading is a repair mechanism that is built into the DNA replicases themselves.  During daughter strand synthesis, an erroneously matched base is excised, and a second attempt at a base pairing is made \cite{Voet}.  Mismatch repair also removes erroneous bases from the daughter strand, but does this shortly after daughter strand synthesis \cite{Voet}.

Two important examples of the second class of repair mechanisms are Nucleotide Excision
Repair (NER) and the SOS response \cite{Voet}.  NER protects a cell from damage due to radiation,
chemical mutagens, and metabolic free radicals by removing damaged portions of the DNA strand and using the other, presumably undamaged strand as a template for re-synthesis of the excised region \cite{Voet}.

The SOS response is a genomic repair mechanism that only activates when there is
extensive damage to the cellular genome.  When DNA damage is sufficiently extensive,
the cell stops growing, and the SOS repair pathways attempt to restore complementarity
to the genome \cite{Voet}.  The SOS response only takes effect when DNA damage is so extensive
that it may be impossible to use undamaged template strands to correctly re-synthesize
damaged portions of the genome.  Thus, although this means that the SOS repair
mechanism is highly error prone, it is evolutionary advantageous for the cell
to repair the genome and risk fixing deleterious mutations, than it is to leave
the damaged genome unrepaired \cite{Voet}.

In recent work with quasispecies models of evolutionary dynamics, quasispecies models 
\cite{QuasReview1, QuasReview2, QuasReview3} considering the first class of repair mechanisms have been studied \cite{MutTann1, MutTann2, MutNowak, MutKess}.  In addition, semiconservative replication, including semiconservative replication with imperfect lesion repair (i.e. not all base-pair mismatches are eliminated), has been considered \cite{SCTann1, SCTann2, SCBrumer1, SCBrumer2}.  Additional effects, such as multiply-gened genomes, as well as multiply chromosomed genomes, have been considered as well \cite{ManyGeneTann, ManyChromTann}.

This paper continues the theme of incorporating various details characteristic of cellular genomes by developing a quasispecies model that takes into consideration the SOS repair mechanism.  The model is highly simplified, and therefore only a first step in developing proper evolutionary dynamics equations with SOS repair.  Nevertheless, because our model is analytically tractable, we believe it is a useful and important initial approach to mathematically modeling the evolutionary aspects of the SOS repair pathway.

\section{Materials and Methods}

\subsection{Definitions and model set-up}

We consider a unicellular population of asexually replicating organisms, whose genomes consist of a single DNA molecule, i.e. one chromosome.  The genome may then be denoted by $ \{\sigma, \sigma'\} $, where $ \sigma $, $ \sigma' $ denote the two strands of the DNA molecule.  If the genome is of length $ L $, then we may write $ \sigma = b_1 \dots b_L $, $ \sigma' = b_1' \dots b_L' $ where each base $ b_i $, $ b_i' $ is chosen from an alphabet of size $ S $ (usually $ = 4 $).  If $ \bar{b}_i $ denotes the base complementary to $ b_i $ (for the standard Watson-Crick bases, the pairings are $ Adenine (A) - Thymine (T) $, $ Guanine (G) - Cytosine (C) $), and $ \bar{\sigma} $ denotes the strand complementary to $ \sigma $, then $ \bar{\sigma} = \bar{b}_L \dots \bar{b}_1 $.  This follows from the antiparallel nature of double-stranded DNA \cite{Voet}.

We let $ n_{\{\sigma, \sigma'\}} $ denote the number of organisms with genome $ \{\sigma, \sigma'\} $, and we assume that replication occurs with a genome-dependent, first-order rate constant, denoted
$ \kappa_{\{\sigma, \sigma'\}} $.  The set of all $ \kappa_{\{\sigma, \sigma'\}} $ defines the {\it fitness landscape}.

The semiconservative replication of the DNA genomes happens in three stages:
\begin{enumerate}
\item Strand separation, whereby each strand of the chromosome separates to act
as a template for daughter strand synthesis.
\item Daughter strand synthesis.  We assume a genome and base-independent mismatch 
probability $ \epsilon $.  This error probability $ \epsilon $ includes all error correction mechanisms, such as proofreading and mismatch repair, that are active during the replication phase of the cell.
\item Lesion repair, where any post-replication mismatches are removed.  Here, there is no longer the parent-daughter strand discrimination that was available during daughter strand synthesis, so in contrast to DNA proofreading and mismatch repair, lesion repair has a $ 50\% $ chance of removing the mutation, and a $ 50\% $ chance of communicating it to the parent strand and fixing the mutation
in the genome.  We also do not assume that lesion repair is perfectly efficient, so that we consider a genome and base-independent probability $ \lambda $ of removing a mismatch.  We call $ \lambda $ the lesion repair efficiency.
\end{enumerate}

In our simplified model, the SOS response is triggered if a given genome has at least $ l_S $ mismatches.  The replication rate of all cells undergoing SOS repair is zero.  We assume that removal of mismatches is catalyzed by an enzyme that binds to a mismatch and then eliminates the mismatch at a rate characterized by a first-order rate constant $ \kappa_{SOS} $.  Therefore, the probability that a given mismatch is eliminated over an infinitesimal time interval $ dt $ is given by $ \kappa_{SOS} dt $. 

In this paper, we will consider the behavior of the model in the limit of infinite sequence length.  If $ \mu \equiv \epsilon L $ is held constant as $ L \rightarrow \infty $, then the probability of an error-free daughter strand synthesis is given by $ (1 - \epsilon)^L \rightarrow e^{-\mu} $.  Therefore, fixing $ \mu $
in the infinite sequence length limit is equivalent to fixing the per-genome replication fidelity.

Finally, we assume that the fitness landscape is defined by a master genome $ \{\sigma_0, \bar{\sigma}_0\} $.  Specifically, we define a genome $ \{\sigma, \sigma'\} $ to be viable, with a first-order growth rate constant $ k > 1 $, if it has fewer than $ l $ mismatches, and if it does not differ from $ \{\sigma_0, \bar{\sigma}_0\} $ by any fixed mutations.  Otherwise, the genome is unviable, with a first-order growth rate constant of $ 1 $.

\begin{figure}
\includegraphics[width = 0.9\linewidth, angle = 0]{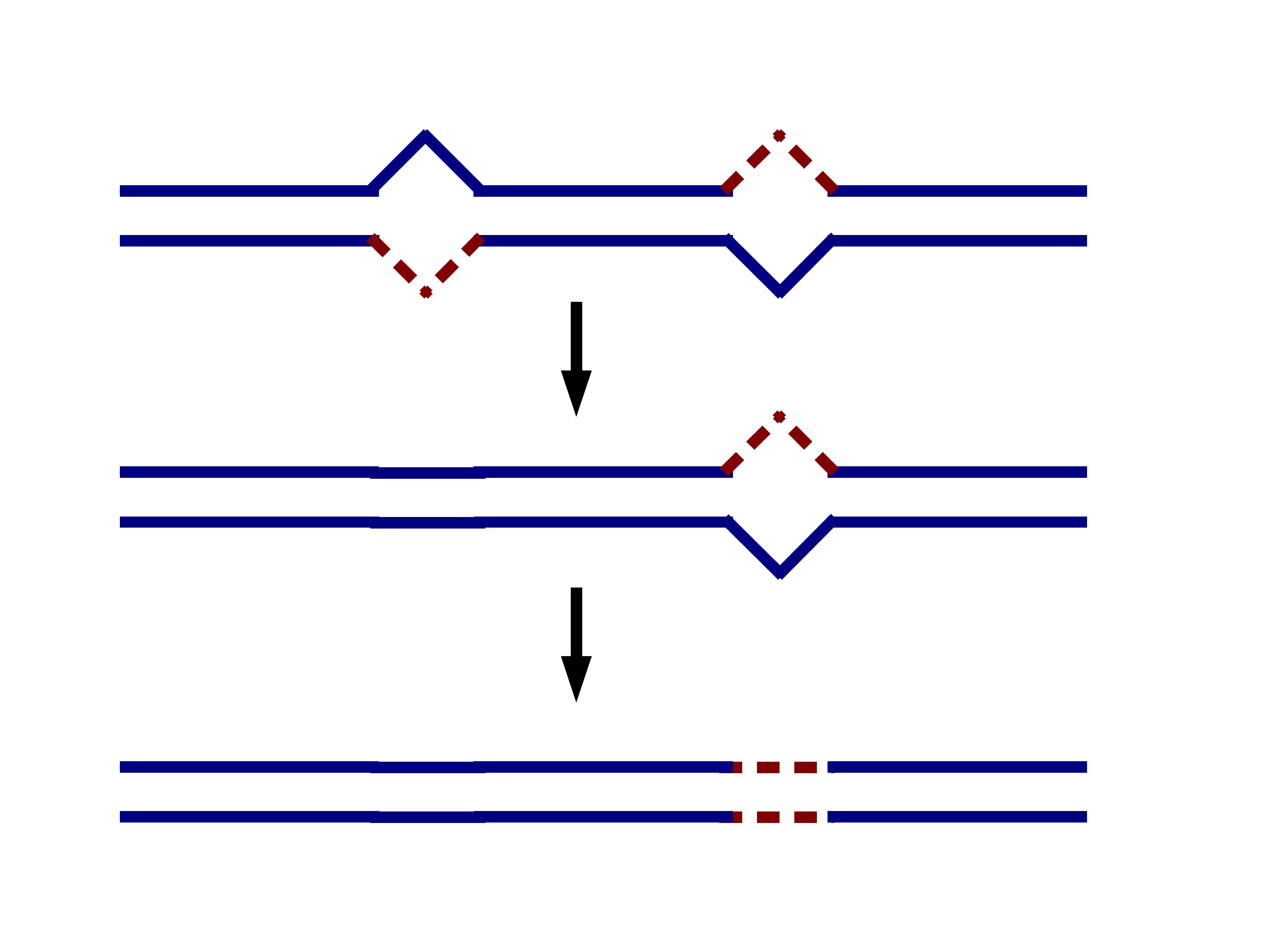}
\caption{Illustration of the SOS repair mechanism being considered in this paper.  A DNA genome with two base-pair mismatches is restored to a fully complementary genome in two repair steps, where during each step a single mismatch (i.e. lesion) is eliminated.  The first lesion is repaired correctly, so that the original base-pair of the master genome strands (solid blue lines) is restored, while the second lesion is repaired incorrectly, so that a mutation (dotted red lines) becomes fixed in the genome.}
\end{figure}

\subsection{Symmetrized population distribution}

We can develop the infinite sequence length equations for our model, assuming an initially prepared clonal population consisting entirely of the genome $ \{\sigma_0, \bar{\sigma}_0\} $.  Because, during replication, only a finite number of mutations are possible, at any time the population will consist of a distribution of genomes $ \{\sigma, \sigma'\} $ where $ \sigma $, $ \sigma' $ differ from either 
$\sigma_0 $ and $ \bar{\sigma}_0 $ in at most a finite number of spots.  Thus, given two gene sequences $ \sigma_1 $, $ \sigma_2 $, if we let $ D_H(\sigma_1, \sigma_2) $ denote the Hamming distance between $ \sigma_1 $ and $ \sigma_2 $ (i.e. the number of sites where $ \sigma_1 $ and $ \sigma_2 $ differ), then either $ D_H(\sigma, \sigma_0) $ and $ D_H(\sigma', \bar{\sigma}_0) $ are finite, or $ D_H(\sigma, \bar{\sigma}_0) $ and $ D_H(\sigma', \sigma_0) $ are finite.

As a result, we can define a strand ordering $ (\sigma, \sigma') $ for a genome $ \{\sigma, \sigma'\} $, where it is understood that $ \sigma $ is a finite Hamming distance from $ \sigma_0 $ and $ \bar{\sigma}_0 $ is a finite Hamming distance from $ \bar{\sigma}_0 $.

A given genome $ (\sigma, \sigma') $ may then be characterized by four parameters $ l_C $, $ l_L $, $ l_R $, and $ l_B $.  We let $ l_C $ denote the number of sites where $ \sigma $ and $ \sigma' $
are both complementary, yet differ from the corresponding bases in $ \sigma_0 $ and $ \bar{\sigma}_0 $.  We let $ l_L $ denote the number of sites where $ \sigma $ differs from $ \sigma_0 $, but $ \sigma' $ is identical to $ \bar{\sigma}_0 $.  We let $ l_R $ denote the number of sites where $ \sigma $ is identical to $ \sigma_0 $, but $ \sigma' $ differs from $ \bar{\sigma}_0 $.  Finally, we let $ l_B $ denote the number of sites where $ \sigma $ and $ \sigma' $ differ from $ \sigma_0 $ and $ \bar{\sigma}_0 $, but are not complementary (for an illustration of these parameters, see \cite{SCTann2, QuasReview3}).

Note that the fitness landscape depends only on $ l_C $, $ l_L $, $ l_R $, and $ l_B $, and hence the fitness of a given organism may be denoted by $ \kappa_{(l_C, l_L, l_R, l_B)} $, where for our
single-fitness-peak landscape we have $ \kappa_{(l_C, l_L, l_R, l_B)} = k $ if $ l_C = 0 $ and $ l_L + l_R + l_B \leq l $, and $ 1 $ otherwise.

By the symmetry of the fitness landscape, and by the symmetry of the initial population distribution, we can group all genomes of identical $ l_C $, $ l_L $, $ l_R $, and $ l_B $, and derive the dynamical equations of the symmetrized population distribution.  We therefore let $ n_{(l_C, l_L, l_R, l_B)} $ denote the total number of organisms in the population whose genomes are characterized by the parameters
$ l_C $, $ l_L $, $ l_R $, and $ l_B $, and we let $ n^{(SOS)}_{(l_C, l_L, l_R, l_B)} $ denote the total number of organisms in the population undergoing the SOS response, whose genomes are similarly
characterized by the parameters $ l_C $, $ l_L $, $ l_R $, and $ l_B $.  The corresponding population fractions are denoted $ z_{(l_C, l_L, l_R, l_B)} $ and $ z^{(SOS)}_{(l_C, l_L, l_R, l_B)} $, respectively.

\subsection{Dynamical equations}

To develop the dynamical equations for both the $ z_{(l_C, l_L, l_R, l_B)} $ and the $ z^{(SOS)}_{(l_C, l_L, l_R, l_B)} $ quantities, we begin by considering a genome $ (\sigma, \sigma') $, characterized by the parameters $ l_C $, $ l_L $, $ l_R $, and $ l_B $.  

We first consider the case where this genome is not undergoing the SOS response.  Then, due to the semiconservative nature of DNA replication, this genome is being destroyed at a rate given by $ -\kappa_{(l_C, l_L, l_R, l_B)} n_{(l_C, l_L, l_R, l_B)} $.  This genome, however, is produced by other
genomes in the population, as a result of replication.  So, consider some other genome $ (\sigma'', \sigma''') $ which produces $ (\sigma, \sigma') $ upon replication.  This can either occur via the $ \sigma'' $ template strand, the $ \sigma''' $ template strand, or both.

If the $ (\sigma'', \sigma''') $ genome is characterized by the parameters $ l_C'' $, $ l_L'' $, $ l_R'' $, and $ l_B'' $, then $ \sigma'' $ differs from $ \sigma_0 $ in $ l_C'' + l_L'' + l_B'' $ spots.  Because sequence lengths are infinite, the probability of a mismatch in one of these spots during daughter strand synthesis is $ 0 $.  In the remaining sites, let $ l_1'' $ denote the number of mismatches that are not
corrected, and $ l_2'' $ denote the number of mismatches that are repaired, but fixed as a mutation in the genome.  Then the resulting genome $ (\sigma, \sigma') $ is characterized by:
\begin{enumerate}
\item $ l_C = l_C'' + l_L'' + l_B'' + l_2'' $
\item $ l_L = 0 $
\item $ l_R = l_1'' $
\item $ l_B = 0 $
\end{enumerate}

The probability of a given set of mutations corresponding to $ l_1'' $, $ l_2'' $, is $ \epsilon^{l_1'' + l_2''} (1 - \lambda)^{l_1''} (\lambda/2)^{l_2''} (1 - \epsilon + \epsilon \lambda/2)^{L - l_C'' - l_L'' - l_B'' - l_1'' - l_2''} $.  The term $ (1 - \epsilon + \epsilon \lambda/2)^{L - l_C'' - l_L'' - l_B'' - l_1'' - l_2''} $ arises
as a probability that the remaining $ L - l_C'' - l_L'' - l_B'' - l_1'' - l_2'' $ sites on $ \sigma'' $ remain identical to $ \sigma_0 $, and the corresponding daughter strand sites are identical to $ \bar{\sigma}_0 $.  The per-site probability of this is the probability of error-free daughter strand synthesis, $ 1 - \epsilon $, plus the probability of a mismatch, times $ \lambda $, the probability that complementarity is restored, times $ 1/2 $, the probability that complementarity is restored correctly.

The degeneracy is given by $ (L - l_C'' - l_L'' - l_B'')!/(l_1''! l_2''! (L - l_C'' - l_L'' - l_B'' - l_1'' - l_2'')!) $,
so in the limit of infinite sequence length the total probability becomes,
\begin{eqnarray}
&   &
\frac{(L - l_C'' - l_L'' - l_B'')!}{l_1''! l_2''! (L - l_C'' - l_L'' - l_B'' - l_1'' - l_2'')!}
\nonumber \\
&   &
\times 
\epsilon^{l_1'' + l_2''} (1 - \lambda)^{l_1''} (\frac{\lambda}{2})^{l_2''}
(1 - (1 - \lambda/2) \epsilon)^{L - l_C'' - l_L'' - l_B'' - l_1'' - l_2''}
\nonumber \\
&   &
\rightarrow 
\frac{1}{l_1''! l_2''!} [\mu (1 - \lambda)]^{l_1''} (\frac{\mu \lambda}{2})^{l_2''}
e^{-(1 - \lambda/2) \mu}
\end{eqnarray}

If $ (\sigma, \sigma') $ is generated by $ \sigma''' $, then we have,
\begin{enumerate}
\item $ l_C = l_C'' + l_R'' + l_B'' + l_2'' $
\item $ l_L = l_1'' $
\item $ l_R = 0 $
\item $ l_B = 0 $
\end{enumerate}
We also obtain an overall transition probability of $ 1/(l_1''! l_2''!) [\mu (1 - \lambda)]^{l_1''} (\mu \lambda/2)^{l_2''} e^{-(1 - \lambda/2) \mu} $.

It is important to note from the $ \sigma'' $ and $ \sigma''' $ results that genomes with $ l_B > 0 $ cannot be generated during replication.  Since SOS repair eliminates mismatches, it follows that a population
where $ l_B $ is initially $ 0 $ for all genomes will always have a population where $ l_B = 0 $.  Therefore, we may assume in subsequent derivations that $ l_B $, $ l_B'' $ are $ 0 $.

Furthermore, note that strands $ \sigma'' $ that are a finite Hamming distance away from $ \sigma_0 $ can only generate daughter genomes where $ l_L = 0 $, while strands $ \sigma''' $ that are a finite
Hamming distance away from $ \bar{\sigma}_0 $ can only generate daughter genomes where $ l_R = 0 $.  Therefore, we may also assume in subsequent derivations that $ l_L $, $ l_R $ are not simultaneously $ > 0 $.

Then for the genomes $ (\sigma, \sigma') $ generated by $ \sigma'' $, we have $ l_C = l_C'' + l_L'' + l_2'' $, and $ l_R = l_1'' $.  Therefore, the restriction on $ (\sigma'', \sigma''') $ is that $ 0 \leq l_2'' \leq l_C $, $ 0 \leq l_L'' \leq l_C - l_2'' $, and $ l_C'' = l_C - l_L'' - l_2'' $.  Note that there is no restriction on $ l_R'' $.

Then for the population number $ n_{(l_C, 0, l_R, 0)} $, we have a contribution from the $ \sigma'' $ strands of
\begin{eqnarray}
&   &
\frac{1}{l_R!} [\mu (1 - \lambda)]^{l_R} e^{-\mu (1 - \lambda/2)}
\nonumber \\
&   &
\times
\sum_{l_2'' = 0}^{l_C} \frac{1}{l_2''!} (\frac{\mu \lambda}{2})^{l_2''}\sum_{l_L'' = 0}^{l_C - l_2''}
\sum_{l_R'' = 0}^{\infty}
\kappa_{(l_C - l_L'' - l_2'', l_L'', l_R'', 0)}
\times \nonumber \\
&  &
n_{(l_C - l_L'' - l_2'', l_L'', l_R'', 0)}
\nonumber \\
\end{eqnarray}
 
A similar expression is obtained for the population number $ n_{(l_C, l_L, 0, 0)} $, except $ l_R $ is replaced with $ l_L $, and the roles of $ l_L'' $ and $ l_R'' $ are exchanged.

It should also be noted that, by the symmetry of the fitness landscape, we have that $ n_{(l_C, l_L, l_R, l_B)} = n_{(l_C, l_R, l_L, l_B)} $.  Another way to note this is that, for a given genome $ (\sigma, \sigma') $, if we change the ordering of the strands so that the first strand is of finite Hamming distance to $ \bar{\sigma}_0 $, and the second strand is of finite Hamming distance to $ \sigma_0 $, then the genome 
$ \{\sigma, \sigma'\} $ must be represented as $ (\sigma', \sigma) $, and is characterized by the parameters $ l_C $, $ l_R $, $ l_L $, and $ l_B $.  If $ \bar{n}_{(l_C, l_L, l_R, l_B)} $ denotes the number of genomes characterized by $ l_C $, $ l_L $, $ l_R $, and $ l_B $, with respect to the $ (\bar{\sigma}_0, \sigma_0) $ strand ordering, then since there is a one-to-one correspondence between genomes 
$ (\sigma, \sigma') $ with parameters $ l_C $, $ l_L $, $ l_R $, $ l_B $ with respect to the first ordering,
and genomes $ (\sigma, \sigma') $ with parameters $ l_C $, $ l_R $, $ l_L $, $ l_B $ with respect to the second ordering, it follows that $ \bar{n}_{(l_C, l_L, l_R, l_B)} = n_{(l_C, l_R, l_L, l_B)} $.  However, since the fitness landscape is invariant under strand ordering, we have $ n_{(l_C, l_L, l_R, l_B)} 
= \bar{n}_{(l_C, l_L, l_R, l_B)} $, so that $ n_{(l_C, l_L, l_R, l_B)} = n_{(l_C, l_R, l_L, l_B)} $.

Taking into consideration the contribution to $ n_{(l_C, 0, 0, 0)} $, we may put everything together and obtain, after changing variables from population numbers to population fractions,
\begin{widetext}
\begin{eqnarray}
&   &
\frac{d z_{(l_C, 0, 0, 0)}}{dt} =
-(\kappa_{(l_C, 0, 0, 0)} + \bar{\kappa}(t)) z_{(l_C, 0, 0, 0)} + \kappa_{SOS} (z^{(SOS)}_{(l_C, 0, 1, 0)} + (1 - \delta_{l_C 0}) z^{(SOS)}_{(l_C - 1, 0, 1, 0)})
\nonumber \\
&   &
+ 2 e^{-\mu (1 - \lambda/2)}
\sum_{l_{1, C} = 0}^{l_C} \sum_{l_1 = 0}^{l_C - l_{1, C}} \sum_{l_2 = 0}^{\infty}
\frac{1}{l_{1, C}!} (\frac{\mu \lambda}{2})^{l_{1, C}} 
\kappa_{(l_C - l_{1, C} - l_1, l_1, l_2, 0)} z_{(l_C - l_{1, C} - l_1, l_1, l_2, 0)}
\nonumber \\
&   &
\frac{d z_{(l_C, 0, l' > 0, 0)}}{dt} =
-(\kappa_{(l_C, 0, l', 0)} + \bar{\kappa}(t)) z_{(l_C, 0, l', 0)}
\nonumber \\
&   &
+ \frac{1}{l'!} [\mu (1 - \lambda)]^{l'} e^{-\mu (1 - \lambda/2)}
\sum_{l_{1, C} = 0}^{l_C} \sum_{l_1 = 0}^{l_C - l_{1, C}} \sum_{l_2 = 0}^{\infty}
\frac{1}{l_{1, C}!} (\frac{\mu \lambda}{2})^{l_{1, C}}
\kappa_{(l_C - l_{1, C} - l_1, l_1, l_2, 0)} z_{(l_C - l_{1, C} - l_1, l_1, l_2, 0)}
\nonumber \\
&   &
\mbox{for $ l' = 1, \dots, l_S - 1 $}
\nonumber \\
&   &
\frac{d z^{(SOS)}_{(l_C, 0, l', 0)}}{dt} =
\kappa_{SOS} [\frac{l' + 1}{2} (z^{(SOS)}_{(l_C, 0, l' + 1, 0)} + (1 - \delta_{l_C 0}) z^{(SOS)}_{(l_C - 1, 0, l' + 1, 0)})
- l' z^{(SOS)}_{(l_C, 0, l', 0)}] - \bar{\kappa}(t) z^{(SOS)}_{(l_C, 0, l', 0)} 
\nonumber \\
&   &
\mbox{for $ l' = 1, \dots, l_S - 1 $}
\nonumber \\
&   &
\frac{d z^{(SOS)}_{(l_C, 0, l' > 0, 0)}}{dt} =
\kappa_{SOS} [\frac{l' + 1}{2} (z^{(SOS)}_{(l_C, 0, l' + 1, 0)} + (1 - \delta_{l_C 0}) z^{(SOS)}_{(l_C - 1, 0, l' + 1, 0)}) -
l' z^{(SOS)}_{(l_C, 0, l', 0)}] - \bar{\kappa}(t) z^{(SOS)}_{(l_C, 0, l', 0)}
\nonumber \\
&   &
+ \frac{1}{l'!} [\mu (1 - \lambda)]^{l'} e^{-\mu (1 - \lambda/2)}
\sum_{l_{1, C} = 0}^{l_C} \sum_{l_1 = 0}^{l_C - l_{1, C}} \sum_{l_2 = 0}^{\infty}
\frac{1}{l_{1, C}!} (\frac{\mu \lambda}{2})^{l_{1, C}}
\kappa_{(l_C - l_{1, C} - l_1, l_1, l_2, 0)} z_{(l_C - l_{1, C} - l_1, l_1, l_2, 0)}
\nonumber \\
&   &
\mbox{for $ l' \geq l_S $}
\end{eqnarray}
\end{widetext}
where $ \bar{\kappa}(t) \equiv \sum_{l_C = 0}^{\infty} \sum_{l_L = 0}^{\infty}
\sum_{l_R = 0}^{\infty} \kappa_{(l_C, l_L, l_R, 0)} z_{(l_C, l_L, l_R, 0)} =
\sum_{l_C = 0}^{\infty} (\kappa_{(l_C, 0, 0, 0)} z_{(l_C, 0, 0, 0)} +
2 \sum_{l' = 1}^{\infty} \kappa_{(l_C, 0, l', 0)} z_{(l_C, 0, l', 0)}) $ is
the mean fitness of the population.

Note that we do not write down the dynamical equations for $ z_{(l_C, l', 0, 0)} $
or $ z^{(SOS)}_{(l_C, l', 0, 0)} $, since they are redundant.

The factor of $ 1/2 $ appearing in the SOS terms arises from the fact that
when a mismatch is removed, it either corrects the daughter strand synthesis
error, or it fixes the mismatch as a mutation in the genome.  In the
former case, the value of $ l_C $ remains unchanged, while in the latter case
it is incremented by $ 1 $.  

It should be noted that this factor is missing in the contribution to $ z_{(l_C, 0, 0, 0)} $ from SOS repair.
The reason for this is that this contribution comes from $ z^{(SOS)}_{(l_C, 0, 1, 0)} $, $ z^{(SOS)}_{(l_C, 1, 0, 0)} $, $ z^{(SOS)}_{(l_C - 1, 0, 1, 0)} $, and $ z^{(SOS)}_{(l_C - 1, 1, 0, 0)} $.  However, because
$ z^{(SOS)}_{(l_C, 0, 1, 0)} = z^{(SOS)}_{(l_C, 1, 0, 0)} $, and $ z^{(SOS)}_{(l_C - 1, 0, 1, 0)} = z^{(SOS)}_{(l_C - 1, 1, 0, 0)} $, we may combine identical terms and eliminate the factor of $ 1/2 $.

The factor of $ l' + 1 $ and $ l' $ in front of the $ \kappa_{SOS} $ rate constant
arises from the fact that the fraction of genomes whose SOS enzymes are bound
to a mismatch is proportional to the total number of mismatches, hence the
resulting SOS rate constant is proportional to the total number of mismatches.

\section{Results and Discussion}

\subsection{Steady-state behavior}

\subsubsection{Definitions and basic equations}

To obtain the steady-state behavior of our model, we begin by introducing some
definitions that will allow us to simplify the calculations.

\begin{enumerate}
\item $ z_1 = z_{(0, 0, 0, 0)} $.
\item $ z_2 = \sum_{l' = 1}^{l} z_{(0, 0, l', 0)} $.
\item $ z_3 = \sum_{l' = l + 1}^{l_S - 1} z_{(0, 0, l', 0)} $.
\item $ z_4 = \sum_{l_C = 1}^{\infty} z_{(l_C, 0, 0, 0)} $.
\item $ z_5 = \sum_{l_C = 1}^{\infty} \sum_{l' = 1}^{l} z_{(l_C, 0, l', 0)} $.
\item $ z_6 = \sum_{l_C = 1}^{\infty} \sum_{l' = l + 1}^{l_S - 1} z_{(l_C, 0, l', 0)} $.
\item $ z^{(SOS)}_{0l'} = z^{(SOS)}_{(0, 0, l', 0)} $.
\item $ z^{(SOS)}_{1l'} = \sum_{l_C = 0}^{\infty} z^{(SOS)}_{(l_C, 0, l', 0)} $.
\item $ z^{(SOS)}_0 = \sum_{l' = 1}^{\infty} z^{(SOS)}_{0l'} $.
\item $ z^{(SOS)} = \sum_{l' = 1}^{\infty} z^{(SOS)}_{1l'} $.
\end{enumerate}
where we set $ l = l_S - 1 $ whenever $ l $ was previously defined as $ \geq l_S $.
The differential equations for $ z_1 $, $ z_2 $, $ z_3 $, $ z_4 $, $ z_5 $, and $ z_6 $ are readily 
derived.  From the equations,
\begin{equation}
\sum_{l_2 = 0}^{\infty} \kappa_{(0, 0, l_2, 0)} z_{(0, 0, l_2, 0)} =
k z_1 + k z_2 + z_3
\end{equation}
and
\begin{eqnarray}
&   &
\sum_{l_C = 0}^{\infty} \sum_{l_{1, C} = 0}^{l_C} 
\sum_{l_1 = 0}^{l_C - l_{1, C}} \sum_{l_2 = 0}^{\infty}
\frac{1}{l_{1, C}!} (\frac{\mu \lambda}{2})^{l_{1, C}} \kappa_{(l_C - l_{1, C} - l_1, l_1, l_2, 0)} 
\times \nonumber \\
&   &
z_{(l_C - l_{1, C} - l_1, l_1, l_2, 0)}
\nonumber \\
&   &
=
e^{\mu \lambda/2} [k z_1 + 2 k z_2 + 2 z_3 + z_4 + 2 z_5 + 2 z_6]
\end{eqnarray}
we obtain,
\begin{eqnarray}
&   &
\frac{d z_1}{dt} = 
-(k + \bar{\kappa}(t)) z_1 + 2 e^{-\mu (1 - \lambda/2)} [k z_1 + k z_2 + z_3] 
\nonumber \\
&   &
+ \kappa_{SOS} z^{(SOS)}_{01}
\nonumber \\
&   &
\frac{d z_2}{dt} =
-(k + \bar{\kappa}(t)) z_2 
\nonumber \\
&   &
+ (f_l(\mu, \lambda) - 1) e^{-\mu (1 - \lambda/2)}
[k z_1 + k z_2 + z_3]
\nonumber \\
&   &
\frac{d z_3}{dt} = 
-(1 + \bar{\kappa}(t)) z_3 
\nonumber \\
&   &
+ (f_{l_S - 1}(\mu, \lambda) - f_l(\mu, \lambda)) e^{-\mu (1 - \lambda/2)}
[k z_1 + k z_2 + z_3]
\nonumber \\
&   &
\frac{d z_4}{dt} = 
-(1 + \bar{\kappa}(t)) z_4 
\nonumber \\
&   &
+ 2 e^{-\mu (1 - \lambda/2)}
[e^{\mu \lambda/2} (k z_1 + 2 k z_2 + 2 z_3 + z_4 + 2 z_5 + 2 z_6) 
\nonumber \\
&   &
- (k z_1 + k z_2 + z_3)] 
+ \kappa_{SOS} [2 z^{(SOS)}_{11} - z^{(SOS)}_{01}]
\nonumber \\
&   &
\frac{d z_5}{dt} = 
-(1 + \bar{\kappa}(t)) z_5 
\nonumber \\
&   &
+ (f_l(\mu, \lambda) - 1) e^{-\mu (1 - \lambda/2)}
\times \nonumber \\
&   &
[e^{\mu \lambda/2} (k z_1 + 2 k z_2 + 2 z_3 + z_4 + 2 z_5 + 2 z_6) 
\nonumber \\
&   &
- (k z_1 + k z_2 + z_3)]
\nonumber \\
&   &
\frac{d z_6}{dt} = 
-(1 + \bar{\kappa}(t)) z_6 
\nonumber \\
&    &
+ (f_{l_S - 1}(\mu, \lambda) - f_l(\mu, \lambda)) e^{-\mu (1 - \lambda/2)}
\times \nonumber \\
&   &
[e^{\mu \lambda/2} (k z_1 + 2 k z_2 + 2 z_3 + z_4 + 2 z_5 + 2 z_6) 
\nonumber \\
&   &
- (k z_1 + k z_2 + z_3)]
\end{eqnarray}
where we define $ f_l(\mu, \lambda) = \sum_{k = 0}^{l} [\mu (1 - \lambda)]^k/k! $ \cite{SCTann2}.
 
We also have,
\begin{eqnarray}
&   &
\frac{d z^{(SOS)}_{0l'}}{dt} = \kappa_{SOS} \frac{l' + 1}{2} z^{(SOS)}_{0 l' + 1} - (l' \kappa_{SOS} + \bar{\kappa}(t)) z^{(SOS)}_{0 l'}
\nonumber \\
&   &
\mbox{for $ l = 1', \dots, l_S - 1 $}
\nonumber \\
&   &
\frac{d z^{(SOS)}_{0l'}}{dt} = \kappa_{SOS} \frac{l' + 1}{2} z^{(SOS)}_{0 l' + 1} - (l' \kappa_{SOS} + \bar{\kappa}(t)) z^{(SOS)}_{0 l'}
\nonumber \\
&   &
+ \frac{1}{l'!} [\mu (1 - \lambda)]^{l'} e^{-\mu (1 - \lambda/2)} [k z_1 + k z_2 + z_3]
\nonumber \\
&   &
\mbox{for $ l' \geq l_S $}
\nonumber \\
&   &
\frac{d z^{(SOS)}_{1l'}}{dt} = \kappa_{SOS} (l' + 1) z^{(SOS)}_{1 l' + 1} - (l' \kappa_{SOS} + \bar{\kappa}(t)) z^{(SOS)}_{1 l'}
\nonumber \\
&   &
\mbox{for $ l' = 1, \dots, l_S - 1 $}
\nonumber \\
&   &
\frac{d z^{(SOS)}_{1l'}}{dt} = \kappa_{SOS} (l' + 1) z^{(SOS)}_{1 l' + 1} - (l' \kappa_{SOS} + \bar{\kappa}(t)) z^{(SOS)}_{1 l'}
\nonumber \\
&   &
+ \frac{1}{l'!} [\mu (1 - \lambda)]^{l'} e^{-\mu (1 - \lambda)} [k z_1 + 2 k z_2 + 2 z_3 + z_4 + 2 z_5 + 2 z_6]
\nonumber \\
&   &
\mbox{for $ l' \geq l_S $}
\end{eqnarray}

We can add these equations to obtain,
\begin{eqnarray}
\frac{d z^{(SOS)}}{dt} 
& = &
-\kappa_{SOS} z^{(SOS)}_{11} - \bar{\kappa}(t) z^{(SOS)}
\nonumber \\
&   &
+ (1 - e^{-\mu (1 - \lambda)} f_{l_S - 1}(\mu, \lambda)) 
\times \nonumber \\
&   &
[k z_1 + 2 k z_2 + 2 z_3 + z_4 + 2 z_5 + 2 z_6] 
\end{eqnarray}

For the purposes of computing the mean fitness at steady-state, we can simplify the system of equations somewhat
by defining $ \tilde{z}_4 = z_4 + 2 z_5 + 2 z_6 $.  We obtain,
\begin{eqnarray}
\frac{d \tilde{z}_4}{dt} 
& = &
-(1 + \bar{\kappa}(t)) \tilde{z}_4 + 2 e^{-\mu (1 - \lambda/2)} f_{l_S - 1} (\mu, \lambda)
\times \nonumber \\ 
&   &
[e^{\mu \lambda/2} (k z_1 + 2 k z_2 + 2 z_3 + \tilde{z}_4) - (k z_1 + k z_2 + z_3)] 
\nonumber \\ 
&   &
+ \kappa_{SOS} [2 z^{(SOS)}_{11} - z^{(SOS)}_{01}]
\end{eqnarray}
For consistency of notation, in what follows we shall simply let $ z_4 $ denote $ \tilde{z}_4 $.

\subsubsection{Determining $ z^{(SOS)}_{01} $, $ z^{(SOS)}_{11} $, and $ z^{(SOS)} $}

To obtain the steady-state behavior of this system of equations, we begin by first solving for the steady-state
of the population undergoing SOS repair.

For $ l' = 1, \dots, l_S - 1 $ we have at steady-state that,
\begin{equation}
z^{(SOS)}_{0 l' + 1} = \frac{2}{l' + 1}(l' + \frac{\bar{\kappa}(t = \infty)}{\kappa_{SOS}}) z^{(SOS)}_{0 l'}
\end{equation}
which gives,
\begin{equation}
z^{(SOS)}_{0 l_S} = \frac{2^{l_S - 1}}{l_S!} [\prod_{l' = 1}^{l_S - 1} (l' + \frac{\bar{\kappa}(t = \infty)}{\kappa_{SOS}})] z^{(SOS)}_{01}
\end{equation}

For $ l' \geq l_S $, we have,
\begin{eqnarray}
z^{(SOS)}_{0 l' + 1} 
& = &
\frac{2}{l' + 1} (l' + \frac{\bar{\kappa}(t = \infty)}{\kappa_{SOS}}) z^{(SOS)}_{0 l'}
\nonumber \\
&   &
- \frac{2}{\kappa_{SOS}} \frac{1}{(l' + 1)!} [\mu (1 - \lambda)]^{l'} 
\times \nonumber \\
&   &
e^{-\mu (1 - \lambda/2)} [k z_1 + k z_2 + z_3]
\end{eqnarray}

This expression has the form of the recursion relation, $ x_{n+1} = a_n x_n - b_n $.  Using mathematical induction, it is possible to prove that $ x_n = a_{n-1} \times \dots \times a_0 x_0 - a_{n-1} \times \dots \times a_1 b_0 - a_{n-1} \times \dots \times a_2 b_1 - \dots - a_{n-1} b_{n-2} - b_{n-1} $.  Therefore,
\begin{eqnarray}
z^{(SOS)}_{0l'} 
& = &
\frac{2^{l' - 1}}{l'!} \sum_{l'' = 1}^{l' - 1} (l'' + \frac{\bar{\kappa}(t = \infty)}{\kappa_{SOS}})
\times \nonumber \\
&   &
[z^{(SOS)}_{01} - \frac{2}{\kappa_{SOS}} e^{-\mu (1 - \lambda/2)} (k z_1 + k z_2 + z_3)
\times \nonumber \\
&   &
\prod_{l'' = 1}^{l_S} \frac{\mu (1 - \lambda)}{2 (l'' + \frac{\bar{\kappa}(t = \infty)}{\kappa_{SOS}})}
\times \nonumber \\
&   &
\sum_{k = 0}^{l' - l_S - 1} \prod_{l'' = 1}^{k} \frac{\mu (1 - \lambda)}{2 (l_S + l'' + \frac{\bar{\kappa}(t = \infty)}{\kappa_{SOS}})}]
\end{eqnarray}
where we define $ \prod_{i = 1}^{0} a_i = 1 $.
 
If we define $ g_{l'}(\mu, \lambda; \bar{\kappa}(t = \infty), \kappa_{SOS}) = 
\prod_{l'' = 1}^{l'} \frac{\mu (1 - \lambda)}{l'' + \frac{\bar{\kappa}(t = \infty)}{\kappa_{SOS}}}
\times \sum_{k = 0}^{\infty} \prod_{l'' = 1}^{k} \frac{\mu (1 - \lambda)}{l' + l'' + \frac{\bar{\kappa}(t = \infty)}{\kappa_{SOS}}} $,
then imposing the requirement that $ \lim_{l' \rightarrow \infty} z^{(SOS)}_{0l'} = 0 $ gives, at steady-state, that,
\begin{eqnarray}
\kappa_{SOS} z^{(SOS)}_{01} 
& = & 
2 e^{-\mu (1 - \lambda/2)} [k z_1 + k z_2 + z_3] 
\times \nonumber \\
&   &
g_{l_S}(\mu/2, \lambda; \bar{\kappa}(t = \infty), \kappa_{SOS})
\end{eqnarray}

Using a similar argument, we obtain,
\begin{eqnarray}
\kappa_{SOS} z^{(SOS)}_{11} 
& = & 
e^{-\mu (1 - \lambda)} [k z_1 + 2 k z_2 + 2 z_3 + z_4] 
\times \nonumber \\
&   &
g_{l_S}(\mu, \lambda; \bar{\kappa}(t = \infty), \kappa_{SOS})
\end{eqnarray}

For the steady-state value of $ z^{(SOS)} $, we have, using the identity
$ \bar{\kappa}(t) = k z_1 + 2 k z_2 + 2 z_3 + z_4 $,
\begin{eqnarray}
&    &
z^{(SOS)} = 
1 - e^{-\mu (1 - \lambda)} 
\times \nonumber \\
&    &
(f_{l_S - 1}(\mu, \lambda) + g_{l_S}(\mu, \lambda; \bar{\kappa}(t = \infty), \kappa_{SOS}))
\nonumber \\
\end{eqnarray}

\subsubsection{Computing $ \bar{\kappa}(t = \infty) $}

Plugging our expressions for $ \kappa_{SOS} z^{(SOS)}_{01} $ and $ \kappa_{SOS} z^{(SOS)}_{11} $ into the steady-state population fractions equations, we obtain,
\begin{eqnarray}
&   &
0 = -(k + \bar{\kappa}(t = \infty)) z_1 
\nonumber \\
&   &
+ 2 e^{-\mu (1 - \lambda/2)} (1 + g_{l_S}(\frac{\mu}{2}, \lambda; \bar{\kappa}(t = \infty), \kappa_{SOS})) [k z_1 + k z_2 + z_3]
\nonumber \\
&   &
0 = -(k + \bar{\kappa}(t = \infty)) z_2 
\nonumber \\
&   &
+ (f_l(\mu, \lambda) - 1) e^{-\mu (1 - \lambda/2)} [k z_1 + k z_2 + z_3]
\nonumber \\
&   &
0 = -(1 + \bar{\kappa}(t = \infty)) z_3 
\nonumber \\
&   &
+ (f_{l_S - 1}(\mu, \lambda) - f_l(\mu, \lambda)) e^{-\mu (1 - \lambda/2)} [k z_1 + k z_2 + z_3]
\nonumber \\
&   &
0 = -(1 + \bar{\kappa}(t = \infty)) z_4 
\nonumber \\
&   &
+ 2 e^{-\mu (1 - \lambda)} (f_{l_S - 1}(\mu, \lambda) + g_{l_S}(\mu, \lambda; \bar{\kappa}(t = \infty), \kappa_{SOS}))
\times \nonumber \\
&   &
[k z_1 + 2 k z_2 + 2 z_3 + z_4]
\nonumber \\
&   &
- 2 e^{-\mu (1 - \lambda/2)} (f_{l_S - 1}(\mu, \lambda) + g_{l_S}(\frac{\mu}{2}, \lambda; \bar{\kappa}(t = \infty), \kappa_{SOS})) 
\times \nonumber \\
&   &
[k z_1 + k z_2 + z_3]
\end{eqnarray}

From these equations we may derive the equality,
\begin{eqnarray}
&   &
k (z_1 + z_2) + z_3 = 
[k (z_1 + z_2) + z_3] e^{-\mu (1 - \lambda/2)} 
\times \nonumber \\
&   &
[k \frac{1 + 2 g_{l_S}(\mu/2, \lambda; \bar{\kappa}(t = \infty), \kappa_{SOS}) + f_l(\mu, \lambda)}
{k + \bar{\kappa}(t = \infty)}
\nonumber \\
&   &
+ \frac{f_{l_S - 1}(\mu, \lambda) - f_l(\mu, \lambda)}{1 + \bar{\kappa}(t = \infty)}]
\end{eqnarray}
Below the error catastrophe, when $ z_1 $, $ z_2 $, $ z_3 $ are not all $ 0 $, we may cancel $ k (z_1 + z_2) + z_3 $ from both sides of the equation and re-arrange to obtain,
\begin{eqnarray}
&   &
\bar{\kappa}(t = \infty)^2 - A(\mu, \lambda; \bar{\kappa}(t = \infty), \kappa_{SOS}) \bar{\kappa}(t = \infty)
\nonumber \\
&  & 
- B(\mu, \lambda; \bar{\kappa}(t = \infty), \kappa_{SOS}) = 0
\end{eqnarray}
where,
\begin{eqnarray}
&   &
A(\mu, \lambda; \bar{\kappa}(t = \infty), \kappa_{SOS}) = 
k [e^{-\mu (1 - \frac{\lambda}{2})} (1 + f_l(\mu, \lambda) 
\nonumber \\
&   & 
+ 2 g_{l_S}(\frac{\mu}{2}, \lambda; \bar{\kappa}(t = \infty), \kappa_{SOS})) - 1] 
\nonumber \\
&   &
+ e^{-\mu (1 - \frac{\lambda}{2})} (f_{l_S - 1}(\mu, \lambda) - f_l(\mu, \lambda)) - 1
\nonumber \\
&   &
B(\mu, \lambda; \bar{\kappa}(t = \infty), \kappa_{SOS}) =
k[e^{-\mu (1 - \frac{\lambda}{2})} (1 + f_{l_S - 1}(\mu, \lambda) 
\nonumber \\
&   &
+ 2 g_{l_S}(\frac{\mu}{2}, \lambda; \bar{\kappa}(t = \infty), \kappa_{SOS})) - 1]
\end{eqnarray}

Beyond the error catastrophe, the mutation rate is sufficiently high that the selective advantage for remaining localized about the $ l_C = 0 $ genomes disappears, so that $ z_1 $, $ z_2 $, and $ z_3 $ drop to $ 0 $.  The relevant steady-state equation is then,
\begin{eqnarray}
0 
& = & 
-(1 + \bar{\kappa}(t = \infty)) z_4 + 2 e^{-\mu (1 - \lambda)} 
\times \nonumber \\
&   &
(f_{l_S - 1}(\mu, \lambda) + g_{l_S}(\mu, \lambda; \bar{\kappa}(t = \infty), \kappa_{SOS})) z_4
\nonumber \\
\end{eqnarray}
which may be solved for $ \bar{\kappa}(t = \infty) $ to give,
\begin{eqnarray}
&   &
\bar{\kappa}(t = \infty) = 
2 e^{-\mu (1 - \lambda)} 
\times \nonumber \\
&   &
[f_{l_S - 1}(\mu, \lambda) + g_{l_S}(\mu, \lambda; \bar{\kappa}(t = \infty), \kappa_{SOS})] - 1
\end{eqnarray}

The error catastrophe occurs at the mutation rate for which the two expressions for the mean equilibrium fitness become equal.

\subsubsection{Limiting Cases}

\noindent
\underline{Case 1:  $ \lambda = 1 $}

\medskip\noindent
When $ \lambda = 1 $, we get for $ l_S > 0 $ that $ g_{l_S}(\mu, \lambda; \bar{\kappa}(t = \infty), \kappa_{SOS}) = 0 $, and that $ f_{l_S - 1}(\mu, \lambda) = 1 $.  Therefore, above the error catastrophe, we obtain $ \bar{\kappa}(t = \infty) = 1 $.  Below the error catastrophe, we have $ A(\mu, 1; \bar{\kappa}(t = \infty), \kappa_{SOS}) = k (2 e^{-\mu/2} - 1) - 1 $, $ B(\mu, 1; \bar{\kappa}(t = \infty), \kappa_{SOS}) = k (2 e^{-\mu/2} - 1) $, giving $ \bar{\kappa}(t = \infty) = k (2 e^{-\mu/2} - 1) $.  These results are in agreement with the solution of the semiconservative quasispecies equations with perfect lesion repair \cite{SCTann1}.

\bigskip\noindent
\underline{Case 2:  $ l_S = \infty $}

\medskip\noindent
When $ l_S = \infty $, then $ g_{l_S}(\mu, \lambda; \bar{\kappa}(t = \infty), \kappa_{SOS}) = 0 $.  Below the error catastrophe, we have $ A(\mu, \lambda; \bar{\kappa}(t = \infty), \kappa_{SOS}) = k [e^{-\mu (1 - \lambda/2)} (1 + f_l(\mu, \lambda)) - 1] - f_l(\mu, \lambda) e^{-\mu (1 - \lambda/2)}
+ e^{-\mu \lambda/2} - 1 $, and $ B(\mu, \lambda; \bar{\kappa}(t = \infty), \kappa_{SOS}) =
k (e^{-\mu (1 - \lambda/2)} + e^{-\mu \lambda/2} - 1) $.  Above the error catastrophe, we have $ \bar{\kappa}(t = \infty) = 1 $.  Both results are in agreement with the semiconservative quasispecies equations with arbitrary lesion repair efficiency \cite{SCTann2}.

\bigskip\noindent
\underline{Case 3:  $ \kappa_{SOS} \rightarrow \infty $}

\medskip\noindent
When $ \kappa_{SOS} \rightarrow \infty $, then $ g_{l_S}(\mu, \lambda; \bar{\kappa}(t = \infty), \kappa_{SOS}) = e^{\mu (1 - \lambda)} - f_{l_S - 1}(\mu, \lambda) $.  Above the error catastrophe, we get that $ \bar{\kappa}(t = \infty) = 1 $.  Below the error catastrophe, we obtain that, $ A(\mu, \lambda; \bar{\kappa}(t = \infty), \kappa_{SOS}) = k[e^{-\mu (1 - \lambda/2)} (1 + f_l(\mu, \lambda) + 2 e^{\mu (1 - \lambda)/2} - 2 f_{l_S - 1}(\mu/2, \lambda)) - 1] + e^{-\mu (1 - \lambda/2)} (f_{l_S - 1}(\mu, \lambda) - f_l(\mu, \lambda)) - 1 $, and $ B(\mu, \lambda; \bar{\kappa}(t = \infty), \kappa_{SOS}) = k[e^{-\mu (1 - \lambda/2)} (1 + f_{l_S - 1}(\mu, \lambda) + 2 e^{\mu (1 - \lambda)/2} - 2 f_{l_S - 1}(\mu/2, \lambda)) - 1] $.  

Taking $ l_S = 1 $ for $ \kappa_{SOS} \rightarrow \infty $ gives $ A(\mu, \lambda; \bar{\kappa}(t = \infty), \kappa_{SOS}) =  k[2 e^{-\mu/2} - 1] - 1 $, and $ B(\mu, \lambda; \bar{\kappa}(t = \infty), \kappa_{SOS}) =
k [2 e^{-\mu/2} - 1] $, so that $ \bar{\kappa}(t = \infty) = k [2 e^{-\mu/2} - 1] $ below the error catastrophe.  This result is identical with the semiconservative quasispecies equations with perfect lesion repair, which makes sense, since here we assume that any lesion is eliminated instantaneously \cite{SCTann2}.

\subsubsection{Optimal Cutoff}

If we assume that $ k >>1 $, and $ \kappa_{SOS} \rightarrow \infty $, then it is possible to find the value of $ l_S $ which maximizes the steady-state mean fitness $ \bar{\kappa}(t = \infty) $.  To do this, we define a normalized mean fitness $ \phi $ to be equal to $ \bar{\kappa}(t = \infty)/k $, and if we divide Eq. (19) by $ k^2 $, we obtain that $ \phi $ is the solution to,
\begin{equation}
\phi^2 - \alpha(\mu, \lambda; \phi, \kappa_{SOS}) \phi - \frac{1}{k} \beta(\mu, \lambda; \phi, \kappa_{SOS}) = 0
\end{equation}
where,
$ \alpha(\mu, \lambda; \phi, \kappa_{SOS}) = e^{-\mu (1 - \lambda/2)} [1 + f_l(\mu, \lambda) + 2 e^{\mu (1 - \lambda)/2} - 2 f_{l_S - 1}(\mu/2, \lambda)] - 1 + \frac{1}{k} [e^{-\mu (1 - \lambda/2)} (f_{l_S - 1}(\mu, \lambda) - f_l(\mu, \lambda)) - 1] $, and $ \beta(\mu, \lambda; \phi, \kappa_{SOS})  = e^{-\mu (1 - \lambda/2)} [1 + f_{l_S - 1}(\mu, \lambda) + 2 e^{\mu (1 - \lambda)/2} - 2 f_{l_S - 1}(\mu/2, \lambda)] - 1 $.

Therefore, for large $ k $ we obtain that $ \phi \rightarrow \lim_{k \rightarrow \infty} \alpha(\mu, \lambda; \phi, \kappa_{SOS}) $, which gives,
\begin{eqnarray}
\phi 
& = & e^{-\mu (1 - \lambda/2)} + 2 e^{-\mu/2} - 1 
\nonumber \\
&   &
+ e^{-\mu (1 - \lambda/2)} (f_l(\mu, \lambda) - 2 f_{l_S - 1}(\mu/2, \lambda))
\end{eqnarray}
so that maximizing $ \phi $ is equivalent to maximizing $ f_l(\mu, \lambda) - 2 f_{l_S - 1}(\mu/2, \lambda) $.

Now, because $ l $ must be re-set to $ l_S - 1 $ whenever we take $ l_S \leq l $, we can only vary $ l_S $ independently of $ l $ whenever $ l_S > l $.  In this regime, the expression $ f_l(\mu, \lambda) - 2 f_{l_S - 1}(\mu/2, \lambda) $ is maximized whenever $ l_S = l + 1 $. 

In the regime where $ l_S \leq l $, $ l $ is re-set to $ l_S - 1 $, and so,
\begin{eqnarray}
 f_l(\mu, \lambda) - 2 f_{l_S - 1}(\mu/2, \lambda) 
 & = &
 f_{l_S - 1}(\mu, \lambda) - 2 f_{l_S - 1}(\mu/2, \lambda)
 \nonumber \\
 & = &
-1 + \mu (1 - \lambda) 
\times \nonumber \\
&   &
\sum_{k = 1}^{l_S - 2} \frac{[\mu (1 - \lambda)]^k}{(k+1)!} (1 - \frac{1}{2^k})
\nonumber \\
\end{eqnarray}
and so this expression is equal to $ -1 $ for $ l_S = 1, 2 $, and then increases with successive values of $ l_S $.    

Now, because $ l $ is re-set to $ l_S -1 $ for $ l_S \leq l $, it follows that we take $ l = l_S - 1 $ for $ l_S \leq l + 1 $.  For $ l = 0 $, we then obtain that $ \phi $ is maximized over $ l_S \leq l + 1 $ for $ l_S = 1 $, while when $ l = 1 $, we obtain that $ \phi $ is maximized over $ l_S \leq l + 1 $ for $ l_S = 1, 2 $.  For $ l \geq 2 $, we obtain that $ \phi $ is maximized over $ l_S \leq l + 1$ for $ l_S = l + 1 $.  

Therefore, in any case, we can maximize $ \phi $ over $ l_S \leq l + 1 $ by taking $ l_S = l + 1 $.  Since we can maximize $ \phi $ over $ l_S \geq l + 1 $ by setting $ l_S = l + 1 $, it follows that $ \phi $ is maximized when $ l_S = l + 1 $.

We reach the conclusion that, when the fitness penalty for having a non-viable genome is sufficiently great, {\it the SOS response will confer a maximum selective advantage if it is activated when and only when the genome has sustained sufficient genetic damage so that it will be unviable without SOS repair}.

\subsection{Stochastic simulations}

We developed stochastic simulations of a unicellular population capable of undergoing the SOS response, in order to numerically test the analytical predictions of our model.  We consider a constant population of genomes that is cycled over every time step.  During each cycle, every genome is allowed to replicate with a probability $ \kappa_{\{\sigma, \sigma'\}} \Delta t $, where $ \kappa_{\{\sigma, \sigma'\}} $ is the first-order growth rate constant of genome $ \{\sigma, \sigma'\} $, and $ \Delta t $ is the length of the time step.  We take $ \Delta t $ to be sufficiently small so that the probability of a given genome replicating more than once during a cycle is negligible. 

We assume that the population initially consists of a clonal population of wild-type (mutation-free) genomes.  The fitness of a given genome $ \{\sigma, \sigma'\} $ is determined by assigning $ l_C, l_L, l_R, l_B $ parameters to the ordered-pairs $ (\sigma, \sigma') $, $ (\sigma', \sigma) $ with respect to the ordered-pair $ (\sigma_0, \bar{\sigma}_0) $.  The fitness is then taken to be the larger of the two fitnesses associated with the two sets of parameters.  

If a genome replicates during a cycle, then it is removed from the population, and the two daughters are added to the population of genomes.  To maintain a constant population size, another, randomly chosen genome is removed from the population as well.

If a daughter genome is produced that has at least $ l_S $ lesions, then it enters the SOS response, and is assigned a replication probability of $ 0 $.  A genome that has initiated the SOS response continues to undergo SOS repair until all lesions have been removed, and a complementary genome has been restored.  During every time step, a genome that is undergoing the SOS response has its lesions scanned, and each lesion is repaired with probability $ \kappa_{SOS} \Delta t $.  In addition to being chosen small enough so that the probability of a given genome replicating more than once during a cycle is negligible, we also choose $ \Delta t $ to be sufficiently small so that the probability that a given genome undergoing the SOS response has more than one lesion repaired during a cycle is also negligible.

The stochastic simulation is allowed to run for a sufficient number of time steps so that the mean fitness of the population does not change significantly, at which point the system is assumed to be at steady-state.

\begin{figure}
\includegraphics[width = 0.9\linewidth, angle = 0]{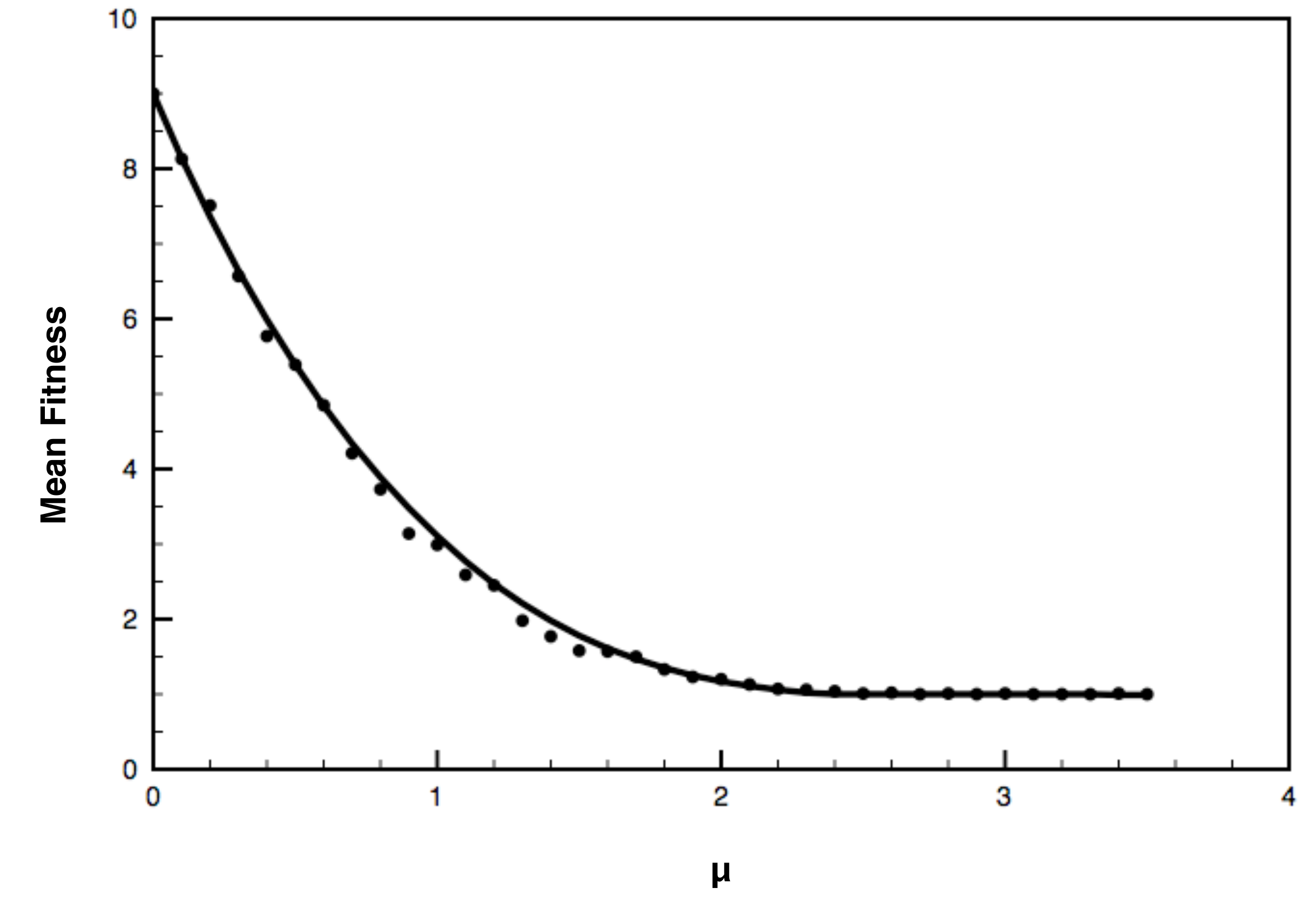}
\caption{Comparison of the mean fitnesses obtained from both stochastic simulations (dots) and the analytical solution (solid line) of our model.  Parameters values are $ k = 9 $, $ l = 4 $, $ l_S = 5 $, $ \lambda = 0.08 $, $ \kappa_{SOS} = 100 $, $ L = 100 $.  The population size was set at $ 1000 $.}
\end{figure}

\begin{figure}
\includegraphics[width = 0.9\linewidth, angle = 0]{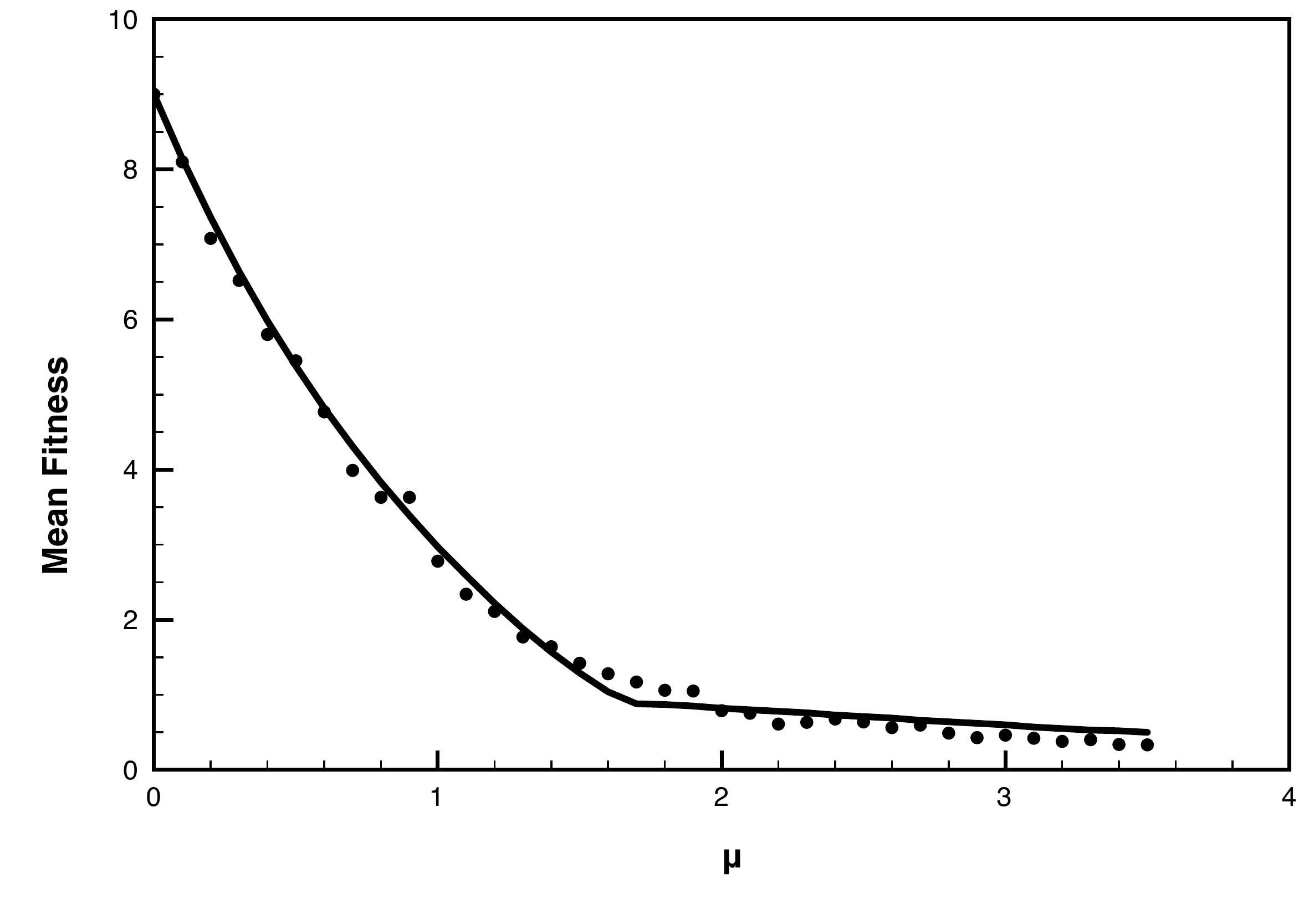}
\caption{Comparison of the mean fitnesses obtained from both stochastic simulations (dots) and the analytical solution (solid line) of our model.  Parameter values are $ k = 9 $, $ l = 4 $, $ l_S = 5 $, $ \lambda = 0.08 $, $ \kappa_{SOS} = 10 $, $ L = 100 $.  The population size was set at $ 1000 $.}
\end{figure}

Figures 2 and 3 show plots comparing the mean fitness obtained from the analytical solution to the mean fitness obtained from the stochastic simulations.  As can be seen from the figures, the agreement between the analytical solution and the stochastic simulation is excellent.

\subsection{Conclusions and Future Research}

This paper developed a quasispecies approach for describing the evolutionary dynamics of a unicellular population that incorporated a simplified model of the SOS response.  The model was a generalization of the single-fitness-peak landscape that is often used in quasispecies theory to study various problems in evolutionary dynamics.  The model was shown to be analytically solvable, and it was found that the solution led to a maximal selective advantage to the SOS response in a manner that is broadly consistent with the behavior of actual organisms.  

For future research, it will be important to move beyond a phenomenological description of the evolutionary dynamics associated with the SOS response, and to consider more realistic models that will allow for quantitative models that can be used in collaboration with experiment.  Nevertheless, as discussed previously, we believe that even this initial model could potentially be used to understand qualitative aspects of the SOS response.  Furthermore, we believe that our model might also be useful for obtaining order-of-magnitude estimates for various parameters associated with the evolutionary dynamics of the SOS response.

\begin{acknowledgments}

This research was supported by the United States - Israel Binational Science Foundation and by the Israel Science Foundation.

\end{acknowledgments}

\end{document}